\documentclass[prd,aps,showpacs,preprintnumbers,amssymb]{revtex4}
\usepackage{graphicx}% Include figure files
\usepackage{dcolumn}% Align table columns on decimal point
\usepackage{bm}% bold math
\begin{document}
\title{
%\hfill\vbox{\hbox{\small PP-04- \qquad\qquad}}\\
%\hfill\vbox{\hbox{\small SU 4252-804 \qquad\qquad}}\\
Fine structure of beta decay endpoint spectrum
}
 \author{Samina S. {\sc Masood}}
 \email{smasood@phy.syr.edu}
 \affiliation{Department of Earth Sciences, SUNY Oswego,
Oswego, NY 13126}
 \author{Salah {\sc Nasri}}
 \email{snasri@physics.umd.edu}
 \affiliation{Department of Physics, University of Maryland, College Park, 
MD 20742-4111,}
\author{Joseph {\sc Schechter}}\email{schechte@phy.syr.edu}
\affiliation{Department of Physics, Syracuse University,
Syracuse, NY 13244-1130}
\date{May 2005}

\begin{abstract}                                                                                             
We note that the fine structure at the endpoint
region of the beta decay spectrum is now essentially
known using neutrino oscillation data, if the mass of one neutrino is 
specified. This may help to 
identify the effects of nonzero neutrino masses in future
experiments. We also give a compact description of the entire range of 
allowed neutrino masses as a function of the third neutrino mass, $m_3$.
A three neutrino assumption is being made.
 An exact treatment
of phase space kinematics is used, in
contrast to the conventional approximate formula. This work is 
independent
of theoretical models; however, additional 
restrictions due 
to the assumption
of a ``complementary ansatz" for the neutrino mass
matrix are also discussed. The ansatz implies that the
values of the three
neutrino masses should approximately be able to form a triangle.
It is noted that most of the presently allowed neutrino mass sets
have this triangular property.
\end{abstract}
\maketitle

\section{Introduction}

     Measurement of the maximum electron energy in beta decay
processes is the original approach \cite{f} to finding the absolute value
of a possible neutrino mass. It is still a subject of great interest    
\cite{aetal} and may yield the first direct measurement of this
crucial quantity. Especially
interesting is the process
  of tritium decay \cite{Mainz, Troitsk},
\begin{equation} 
{}^3H \rightarrow {}^3He^+ +e^- +{\bar \nu}_e .
\label{tritiumdecay}
\end{equation}  
    In this paper, we discuss how results 
obtained from present neutrino oscillation experiments \cite{nuosc}
yield characteristic shapes for the
beta decay endpoint spectrum
which could help in signal identification \cite{morerefs}.
The main point is that the present oscillation data already
predict fairly reliably a bi-unique description of the shape
of the endpoint spectrum if the mass of any neutrino
is specified. This fortunate situation arises since, as we will
discuss, the only unknown mixing angle is strongly bounded and hence 
cannot have much effect on the endpoint spectrum shape.
 So, running over the possible values of
 the third neutrino mass $m_3$, we would get a ``catalog"
of shapes which can be compared with the 
experimental shape to find a best fit. Of course, there
are a number of practical corrections to the
 observed end point spectrum other than just the phase
space and neutrino mixing effects to be considered here.
These include \cite{bv} (i) Different final masses of 
${}^3He^+$ due to different final atomic electron states,
(ii) Corrections due to binding of ${}^3H$ in a molecule,
(iii) Resolution function of the detector, (iv) Final state effects. 
The corrections should be made on each ``page" of the catalog
obtained.

  We start in section II with
an exact numerical treatment of the phase space,
which contains a relevant correction to the
approximate treatment often used. In section III
we show that, as $m_3$ decreases from
about the highest value considered
to be consistent with information from cosmology, there are,
in general, two solutions:
type I where $m_3$ is the largest of the neutrino masses
and type II where  $m_3$ is the smallest of all the neutrino masses.
Below about $m_3=0.052$ eV only the type II solutions are allowed.
A characteristic feature is that the neutrinos 1 and 2
are extremely close in mass; their splittings range from
about a ten thousandth of an eV to a hundredth of an eV.  
The end point spectrum shapes, showing the slope discontinuities
corresponding to the vanishing of each neutrino's contribution,
are plotted to illustrate these features. With a very great accuracy
that might be achieved in the future, the spacings of the neutrinos
could conceivably survive the needed corrections mentioned above.
However the most practical procedure would probably be to integrate
each corrected predicted spectrum shape (of definite $m_3$ and type)
 over an energy interval and then look to see which one is best fit by the
experimental value.

 In section IV we discuss some model
restrictions on the ranges of $m_3$ which
lead to type I and type II solutions. These restrictions arise by assuming  
a so-called ``complementary ansatz" for
the neutrino mass matrix. The observation of $m_3$                                     
in the non-allowed ranges would constitute a falsification of the 
ansatz.

    Finally, section V contains a brief summary and some discussion.

\section{Phase space kinematics}

   Taking the tritium decay example and assuming one massive
neutrino for the time being, let $M$ be the mass of the tritium
atom, $M'$ the mass of ${}^3He^+$, $m_e$ the electron mass and
$m_\nu$ the neutrino mass. Often \cite{k}, the kinetic part of the 
recoil nucleus energy is neglected so one has an easy
 approximate formula
for the maximum electron energy:
\begin{equation}
E_{max}(approx)=M-M'-m_\nu.
\label{appform}
\end{equation}
The exact formula, corresponding to the physical situation where
the neutrino and the recoil nucleus both emerge
 with the same velocity is 
\begin{equation}
E_{max}=\frac{1}{2M}[M^2+m_e^2-(m_\nu+M')^2].
\label{exactform}
\end{equation}

To get an idea of the accuracy of the approximate formula
we use \cite{www} the input masses (in MeV):
\begin{equation}
M= 2809.431935572 \hskip2cm M'=2808.902399642,
\label{decimals}
\end{equation}
to obtain Table I. Notice that in Eq. (\ref{decimals})
we have, for the purpose of
conveniently ``tracking" digits in
 this illustration, included more digits than 
warranted for the experimental accuracy of $M$ and $M'$.
\begin{table}[htbp]
\begin{center}
\begin{tabular}{ccc}
\hline \hline $m_\nu$(eV)&$E_{max}$(MeV)&$\delta E_{max}$(eV)
 \\
\hline                                               
\hline
10&0.5295225&3.431
\\
1&0.529531497&3.433
\\
0.1&0.529532397&3.433
\\
0.01&0.529532487&3.433
\\
0.001&0.529532496&3.433 \\
\hline
\hline
\end{tabular}
\end{center}                                                          
\caption[]{Comparison of exact and approximate
 maximum electron energy
for different neutrino masses. Note that
$\delta E_{max}= E_{max}(approx)-E_{max}$.}
\label{compareEmax}
\end{table}                                             

   In the second column of Table \ref{compareEmax},
as one goes to lower neutrino masses
the change in the maximum electron energy occurs, as expected, in
the decimal place corresponding to the neutrino mass. The third
column shows that there is a
 correction to the approximate formula Eq.(\ref{appform})
of 3.43 eV which is essentially independent of neutrino mass.
This value is considerably larger than (as we will review) the
still allowed values of neutrino mass and hence is very relevant.
Its analytic form may be obtained by subtracting Eq.(\ref{exactform})
from  Eq.(\ref{appform}) and then neglecting the neutrino mass
dependence:
\begin{equation}
\delta E_{max}= 
E_{max}(approx)-E_{max}\approx \frac{1}{2M}[(M-M')^2-m_e^2].
\label{deltaEmax}
\end{equation}
Since one gets a decent approximation to $E_{max}$ by 
subtracting the (neutrino mass independent) expression for
$\delta E_{max}$ just given
 from $E_{max}(approx)$, the difference of any two
maximum electron energies essentially equals minus the difference of the
corresponding neutrino masses.

\begin{figure}[htbp]
%\begin{center}
%\epsfxsize = 12cm
%\ \epsfbox{test1a.eps}
%\mbox{\epsfig{file=test1a.eps,height=4in,width=4in,angle=0}}
%\end{center}
\centering
\rotatebox{270}
{\includegraphics[width=5cm,height=10cm,clip=true]{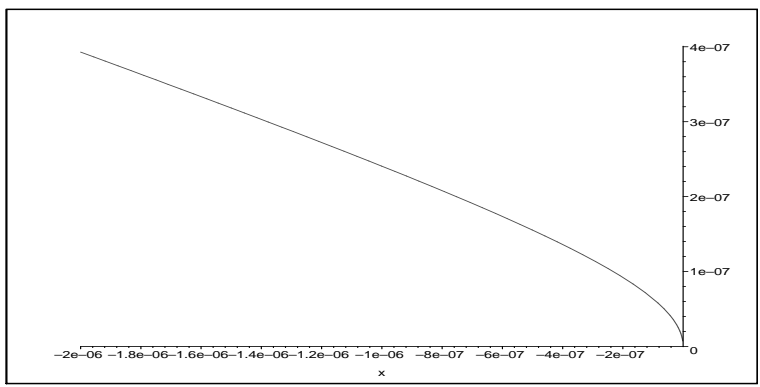}}
\caption[]{Plot of the phase space factor $\Phi$ for a 1 eV neutrino.
 $x= E-Emax$ has units of MeV while $\Phi$ has units of $MeV^2$}
\label{1eV}
\end{figure}

   Considering that there is a relevant correction to the approximate
end point formula, it seems 
prudent to also calculate the phase space factor 
exactly. This does not seem to have been used before but 
is straightforward to get by standard methods \cite{rpp}. 
We write the distribution in the electron energy, $E$ as, 
\begin{equation}
|\frac{d\Gamma}{dE}|=\frac{1}{(2\pi)^3} \frac{|{\cal M}|^2}{(2M)^2} 
\Phi(E),
\label{Edistrib}
\end{equation}
where $|{\cal M}|^2$ stands for a suitably spin averaged squared
amplitude and $\Phi(E)$ is the desired phase space factor. 
After integration over one coordinate of the Dalitz
diagram, we find:
\begin{eqnarray}
\Phi(E)=\sqrt{(E_2^{*2}-M'^2)(E_3^{*2}-m_e^2)},
\nonumber \\
E_2^*=\frac{m_{12}^2-m_{\nu}^2+M'^2}{2m_{12}},
\nonumber \\
E_3^*=\frac{M^2-m_{12}^2-m_e^2}{2m_{12}},
\nonumber \\
m_{12}=\sqrt{M^2+m_e^2-2ME}.
\label{phasespace}
\end{eqnarray}
In carrying out the integration we assumed that there
is a negligible energy dependence introduced
from $|{\cal M}|^2$. This is reasonable since we are only
interested in the very small endpoint region. The shape
of $\Phi$ in the endpoint region is illustrated in Fig.
\ref{1eV} for the case of a one electron volt neutrino.

\section{Case of three neutrinos}

    In the realistic three neutrino case, the ordinary beta
decays actually correspond to decays
with different weights into three neutrinos of different 
mass. The effective phase space factor is:
\begin{equation}
\Phi_{eff}(E) = \sum_{i=1}^{3}|K_{1i}|^2\Phi_i(E)\theta(E_{max,i}-E),
\label{effphspace}
\end{equation}
where $\Phi_i(E)$ is the phase space factor 
 and $E_{max,i}$  the maximum electron energy
for a neutrino of
mass $m_i$. The $K_{1i}$ are matrix elements of the lepton mixing 
matrix, displayed for convenience in the Appendix. Notice that no CP 
phases contribute.  

    As the result of a number of beautiful experiments \cite{nuosc}, there
is now a great deal of information available about the squared mass
differences and mixing angles involved in neutrino physics.
Essentially, assuming that the three neutrino scenario 
is correct, if the absolute mass of any one neutrino is specified,
everything in Eq. (\ref{effphspace}) up to a possible
two-fold ambiguity is known with useful accuracy.
 According to a recent
analysis \cite{mstv} it is possible to extract from the data to good 
accuracy,
two squared neutrino mass differences: $m_2^2-m_1^2$ and $|m_3^2-m_2^2|$,
and two inter-generational mixing angle squared sines: $s_{12}^2$
 and $s_{23}^2$.
Furthermore the inter-generational mixing parameter $s_{13}^2$ is found to
be very small. Specifically, we will adopt,
\begin{eqnarray}
 A \equiv m_2^2-m_1^2 &=& 6.9 \times 10^{-5} eV^2, \nonumber \\
 B \equiv |m_3^2-m_2^2| &=& 2.6 \times 10^{-3} eV^2.
\label{massdifferences}
\end{eqnarray}
The uncertainty in these determinations is roughly $25 \%$.
Similarly for definiteness  we will adopt the best fit values for
$s_{12}^2$
and $s_{23}^2$ obtained in the same analysis:
\begin{equation}   
s_{12}^2 = 0.30,\quad  s_{23}^2 = 0.50.
\label{exptmixingangles}
\end{equation}
 These mixing parameters also have about a $25 \%$ uncertainty.
The experimental value of $s_{13}^2$ was less accurately
determined. At present
only the  3 $\sigma$ bound,
\begin{equation}
s_{13}^2 \leq 0.047,
\label{onethreebound}
\end{equation}
is available and we will choose, for definiteness,
 $s_{13}^2=0.04$.
Then, our values for the weighting coefficients in
Eq. (\ref{effphspace}) will be taken as,
\begin{eqnarray}
|K_{11}|^2&=&(c_{12}c_{13})^2=0.672,
\nonumber \\
|K_{12}|^2&=& (s_{12}c_{13})^2=0.288,
\nonumber \\
|K_{13}|^2&=& (s_{13})^2=0.040.
\label{weights}
\end{eqnarray}                                                                
Future improvements in these factors are not
expected to make qualitative changes in the
endpoint spectrum shapes.  In particular, even though the 
exact value of $|K_{13}|^2$ is not known, the bound of Eq.(\ref{onethreebound})
guarantees that the contribution of neutrino three
 will be no larger than the amount to be given.

    The goal would be to fit the endpoint spectrum shape
observed in a future tritium decay experiment to one of
 the family of shapes that we can now find by running through
the possible values of the third neutrino mass, $m_3$.
 For a given value of $m_3$
 one can obtain from Eqs.(\ref{massdifferences}) two different
solutions for the other
masses $m_1$ and $m_2$. We call the solution where $m_3$ is the largest 
neutrino
mass, the type I case. The case where $m_3$ is the smallest neutrino mass 
is
designated type II. $m_1$ will be determined from the assumed value of
$m_3$ as,
\begin{equation}
m_1^2=m_3^2-A \mp B,
\label{findmone}
\end{equation}
where the upper and lower sign choices respectively refer
to the type I and type II cases. In either case
we find $m_2$ as,
\begin{equation}
m_2^2=A+m_1^2.
\label{findmtwo}
\end{equation}                                                         
Which values of $m_3$ are allowed? A recent cosmology bound
 \cite{cosmobound} based on structure formation requires,
\begin{equation}
m_1+m_2+m_3<0.7 \textrm{ eV}. 
\label{numassbound}
\end{equation}
Thus values of $m_3$ greater than about
0.3 eV are physically disfavored.
However, it seems extremely important that this bound be verified
independently by laboratory experiments like tritium beta decay.
At the lower end, we see from Eq.(\ref{findmone})
that the type I solutions must satisfy,
\begin{equation}
m_3>\sqrt{A+B} \approx 0.0517 \textrm{eV}.
\label{typeIbound}
\end{equation}
On the other hand, the type II solutions need only
obey $m_3>0$ at the lower end. Some typical 
allowed solutions are shown in Table \ref{typicalsolutions}.   

\begin{table}[htbp]
\begin{center}
\begin{tabular}{cccc}
\hline \hline type & $m_1$(eV)&$m_2$(eV)&$m_3$(eV)  
 \\
\hline
\hline
I & 0.2955& 0.2956& 0.3 
\\
II & 0.3042& 0.3043& 0.3   
\\
I & 0.0856& 0.0860& 0.1 
\\
II & 0.1119& 0.1123& 0.1 
\\
I & 0.0305& 0.0316& 0.06 
\\
II & 0.0783& 0.0787& 0.06 
\\
I & 0.0000& 0.0083& 0.0517
\\
II & 0.0643& 0.0648& 0.04 
  \\
II & 0.0541& 0.0548& 0.02 
 \\
II & 0.0506& 0.0512& 0.005         
  \\
II & 0.0503& 0.0510& 0.001 
 \\
\hline
\hline
\end{tabular}
\end{center}
\caption[]{Typical solutions for $(m_1, m_2)$
as $m_3$ is lowered from about the
highest value which is experimentally reasonable. 
 In the type I
solutions $m_3$ is the largest mass
while in the type II solutions $m_3$ is the smallest mass.}
\label{typicalsolutions}
\end{table}                                                     

 This table contains a great deal of information
 in a very small space. For orientation, we remark that the type I
situation is usually called the "normal hierarchy" while the type II
situation is usually called the "inverted hierarchy". Actually the table shows
that over a large portion of the currently favored neutrino mass range
(i.e. roughly from $m_3$ =0.1 to 0.3 eV as opposed to $m_3$ = 0 to 0.1 eV) 
and for either type I or type II, the neutrino mass spectrum
 is better described as an almost degenerate situation rather than one
involving a hierarchy. If one goes above $m_3$ = 0.3 eV
the degeneracy is even enhanced. Below about $m_3$ = 0.1
hierarchies can exist. Note that the  maximum ratio associated with a normal
hierarchy corresponds to
 $m_3/m_2$  only about 6 and is achieved just when (while decreasing $m_3$)
this type vanishes. On the other hand, we see that the ratio $m_2/m_3$ can become
indefinitely large (as $m_3$ goes to zero) for the inverted hierarchy case.
 Another interesting   
feature shown in this table is the near
mass degeneracy of neutrinos one and two for all the solutions. 
The table  precisely shows how this near degeneracy gets somewhat
 relaxed as the mass $m_3$ decreases from the top allowed value.
Finally, the table illustrates
 the vanishing of the type I solutions,
 consistently with Eq. (\ref{typeIbound}),
as $m_3$ is decreased.  

\begin{figure}[htbp]
\centering
\rotatebox{270}
{\includegraphics[width=5cm,height=10cm,clip=true]{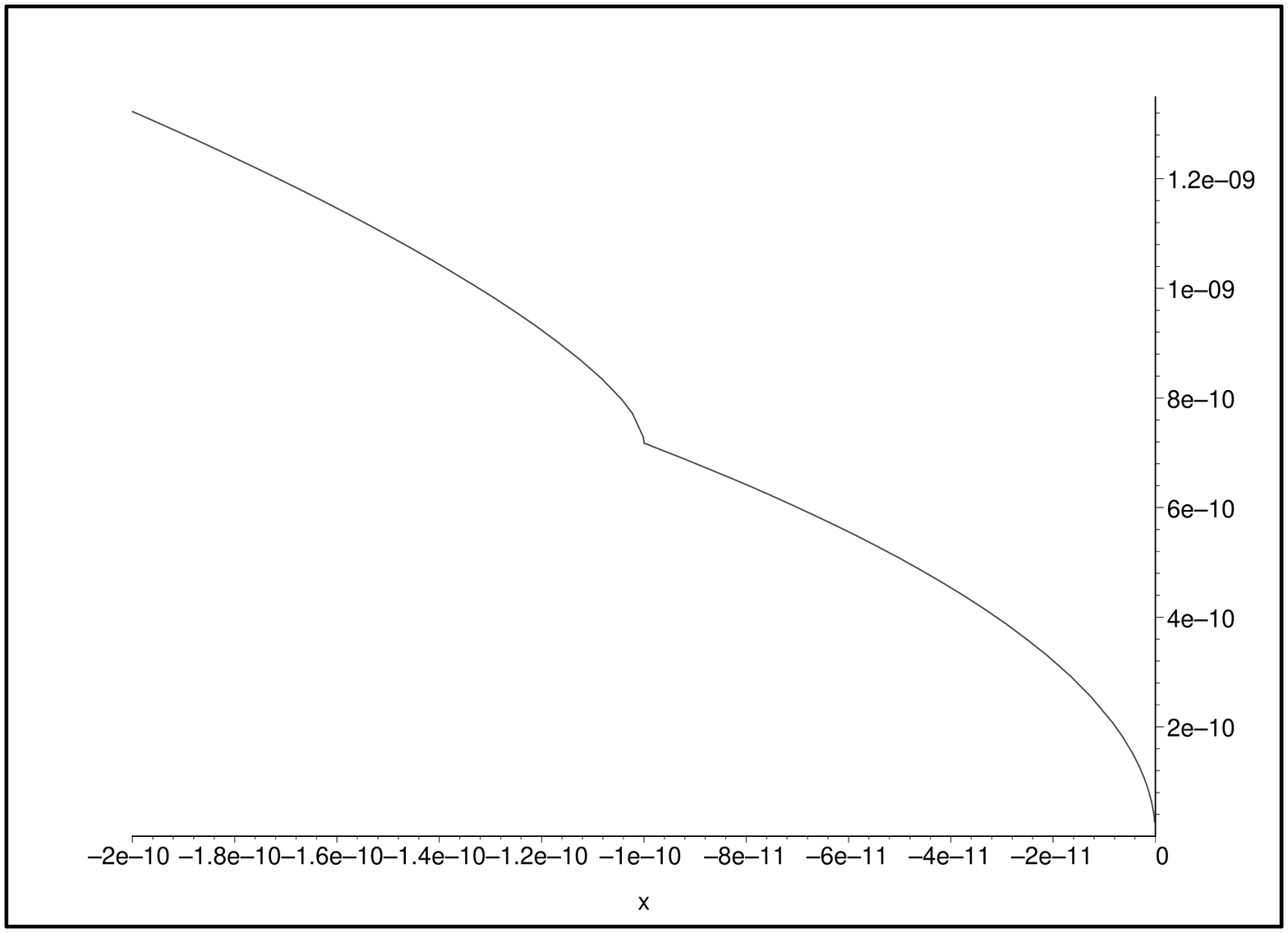}}
\caption[]{Plot of $\Phi_{eff}$ for $m_3=0.3 eV$ and $m_1,m_2=
0.2955,0.2956 eV$ in the region showing the $m_1,m_2$ structure.
 $x= E-E_{max,1}$ has units of MeV while $\Phi_{eff}$ has units of 
$MeV^2$}
\label{I-0.3}
\end{figure}

\begin{figure}[htbp]
\centering
\rotatebox{270}
{\includegraphics[width=5cm,height=10cm,clip=true]{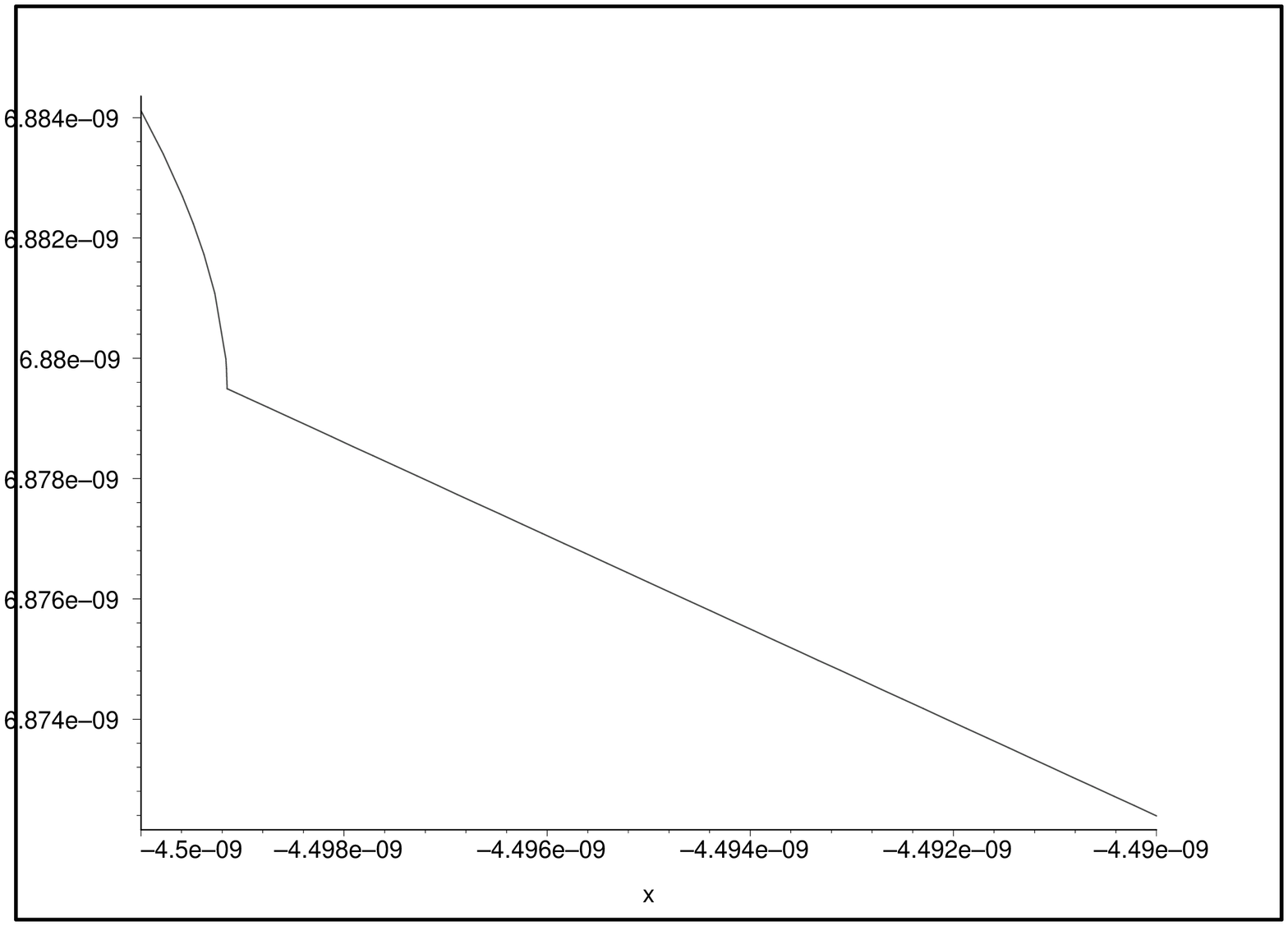}}
\caption[]{Plot of $\Phi_{eff}$ for $m_3=0.3 eV$ and $m_1,m_2=
0.2955,0.2956 eV$ in the region showing the $m_3$ structure.
 $x= E-E_{max,1}$ has units of MeV while $\Phi_{eff}$ has units of
$MeV^2$}
\label{I-0.3extra}
\end{figure}

    Now let us examine the 
endpoint spectrum shape computed from Eq. (\ref{effphspace}).
Figure \ref{I-0.3} shows the endpoint spectrum in the type I
case where $m_3=0.3$ eV and correspondingly $m_1,m_2=0.2955,0.2956$ eV.
In addition to the vanishing of $\Phi_{eff}$
 at $x=E-E_{max,1}=0$ due
to neutrino one there is also a nearby discontinuity of slope
at $x=E_{max,2}-E_{max,1}$ due to neutrino two. The presence
of two distinguishing features may be easier to recognize than just one
alone. However, the difference between these two points is read off
to be about one ten-thousandth of an eV (which, as remarked
previously, also is the difference between the two
neutrino masses seen in Table \ref{typicalsolutions}).
This seems to be a rather small number for experimental detection. 
 The slope discontinuity due to neutrino three is still further
to the left and is shown, greatly magnified, in Fig. \ref{I-0.3extra}.
It is only about five one thousandths of an eV
away from the others. However the signal for neutrino three
is suppressed by the small value of $s_{13}^2$, as discussed above.
 Conceivably,
requiring the presence of three distinguished points
together may make
the neutrino mass signal somewhat easier to recognize in the future. 

\newpage

\begin{figure}[htbp]
\centering
\rotatebox{270}
{\includegraphics[width=5cm,height=10cm,clip=true]{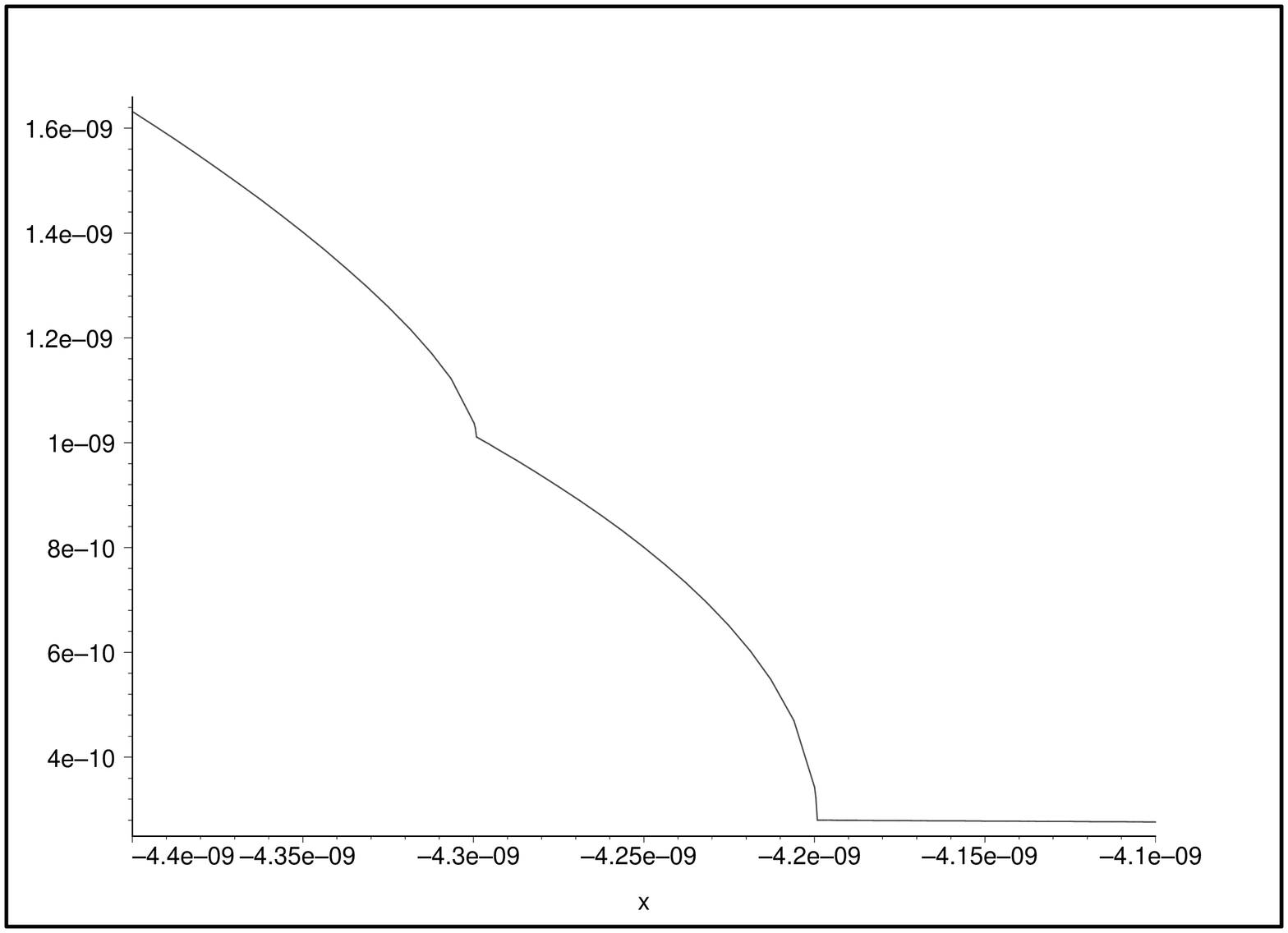}}
\caption[]{Plot of $\Phi_{eff}$ for $m_3=0.3 eV$ and $m_1,m_2=
0.3042,0.3043 eV$ in the region showing the $m_1,m_2$ structure.
 $x= E-E_{max,3}$ has units of MeV while $\Phi_{eff}$ has units of
$MeV^2$}
\label{II-0.3}
\end{figure}

\begin{figure}[htbp]
\centering
\rotatebox{270}
{\includegraphics[width=5cm,height=10cm,clip=true]{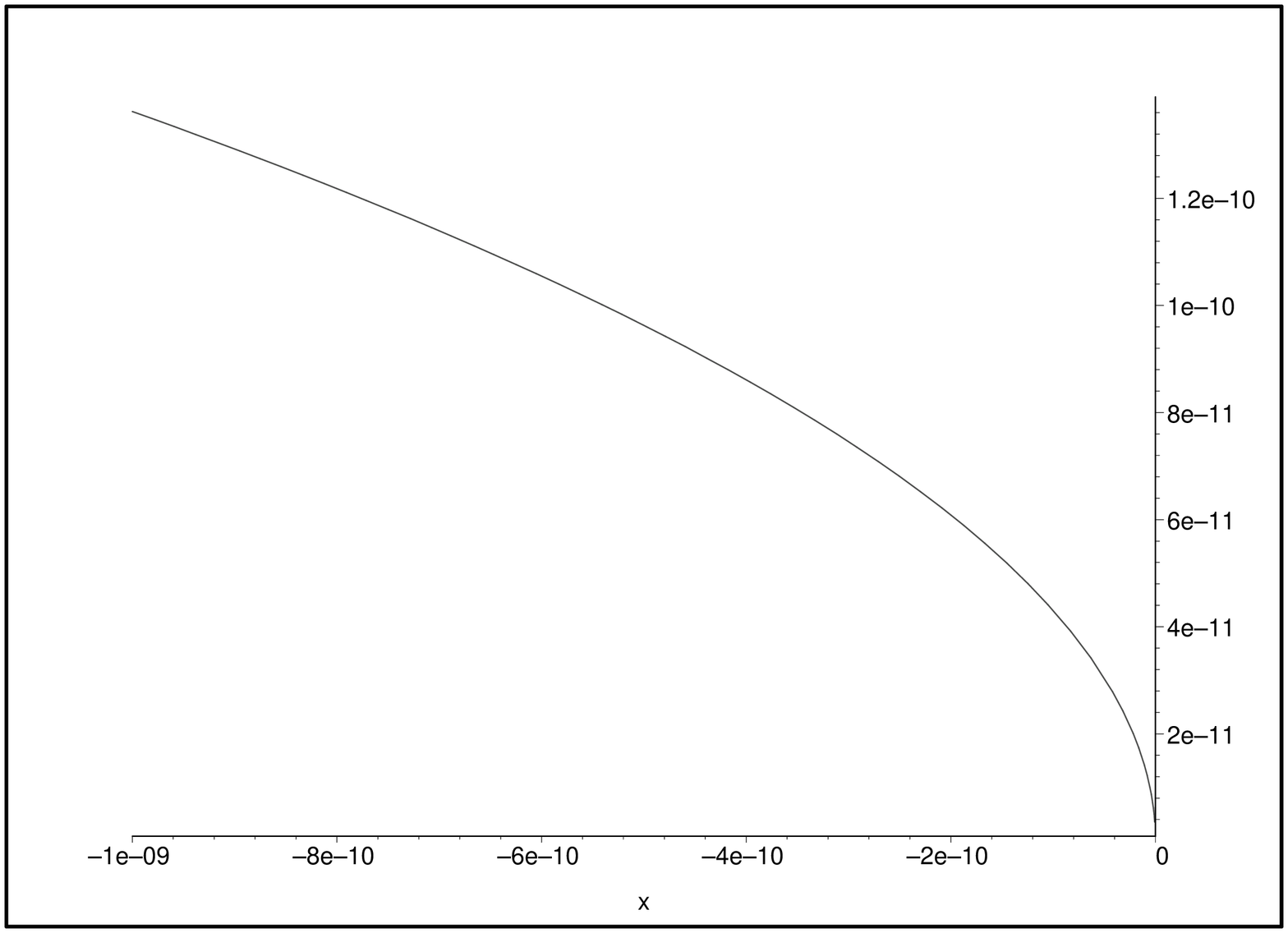}}
\caption[]{Plot of $\Phi_{eff}$ for $m_3=0.3 eV$ and $m_1,m_2=
0.3042,0.3043 eV$ in the region showing the $m_3$ structure.
 $x= E-E_{max,3}$ has units of MeV while $\Phi_{eff}$ has units of
$MeV^2$}
\label{II-0.3extra}
\end{figure}

    The endpoint spectrum for the type II solution with $m_3=0.3$ eV 
is similarly plotted in Figs. \ref{II-0.3} and \ref{II-0.3extra}.
For this case, since $m_3$ is the smallest mass, one recognizes
that the curve in Fig. \ref{II-0.3} does not quite vanish at
the right end (larger electron energy) but goes to a small
value controlled by the small value of $s_{13}$.  Clearly,
as discussed above, finding a more accurate value of $K_{13}$ consistent with the experimental
bound will have a small effect on the endpoint spectrum shape. The actual
vanishing at the right end is shown, magnified, in Fig. \ref{II-0.3extra}.
The spacings between the various neutrino points are essentially
the same in magnitude as in the type I case. However the type I
and II cases can, in principle, be distinguished by noting that
the long interval comes first in the type II case.
 
\newpage
                                                        
\begin{figure}[htbp]
\centering
\rotatebox{270}
{\includegraphics[width=5cm,height=10cm,clip=true]{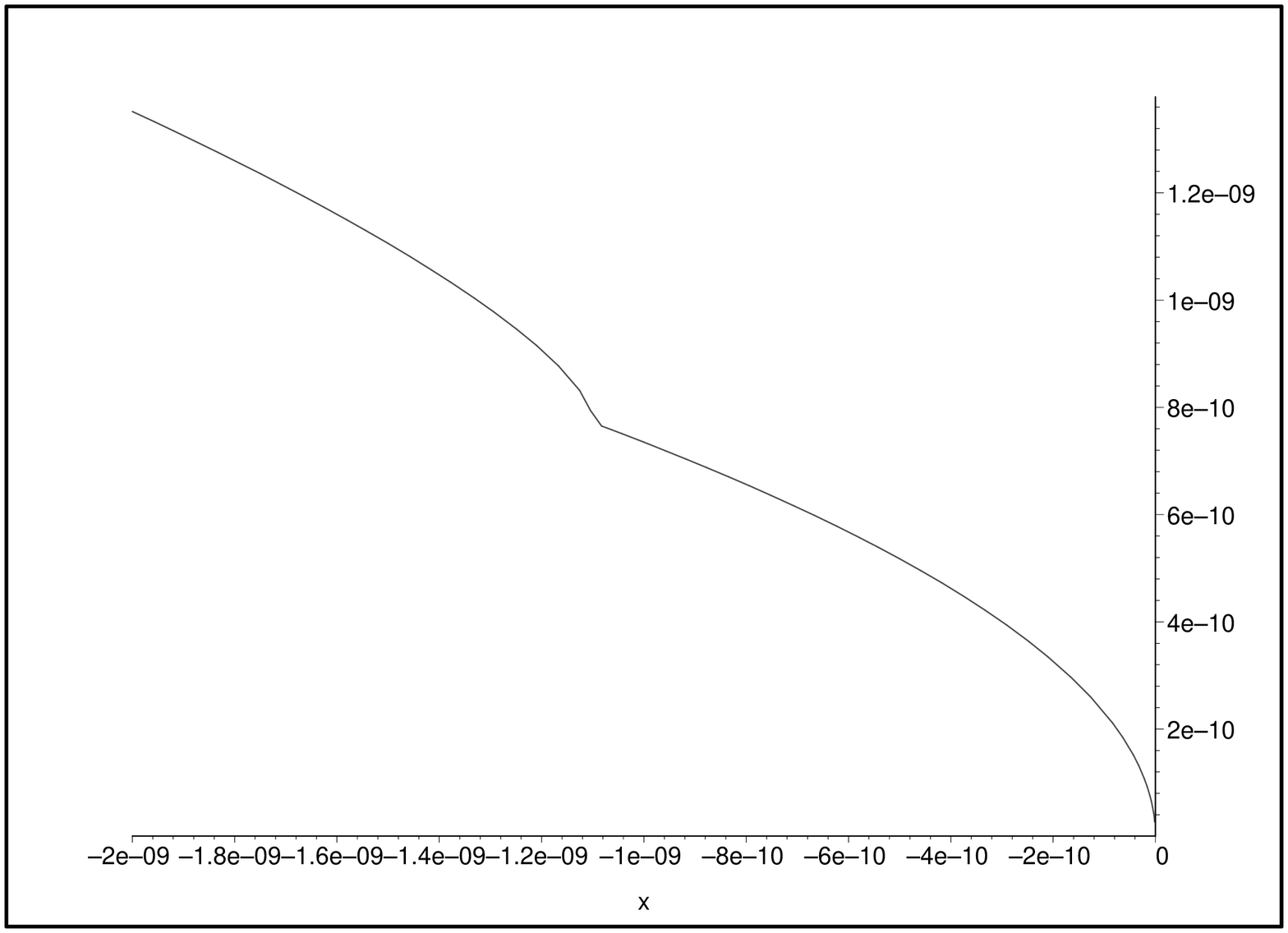}}
\caption[]{Plot of $\Phi_{eff}$ for $m_3=0.06 eV$ and $m_1,m_2=
0.0305,0.0316 eV$ in the region showing the $m_1,m_2$ structure.
 $x= E-E_{max,1}$ has units of MeV while $\Phi_{eff}$ has units of
$MeV^2$}
\label{I-0.06}
\end{figure}

    The figures for lower values of $m_3$ are very similar
in shape. We give one more plot here
for comparison. As $m_3$ 
decreases,
the distinguishing points for neutrinos one and two move 
somewhat apart. This can be understood since the corresponding
masses are seen in Table \ref{typicalsolutions}
to move apart and, according to Eqs. 
(\ref{appform}) and 
(\ref{deltaEmax}) the splitting of the maximum electron
energies is, to a very decent approximation, the same as the 
neutrino mass splitting. In Fig. \ref{I-0.06} the type I case
with $m_3=0.06$ eV is shown. Here the splitting between
the neutrino one and two points is one order of magnitude larger
than for the corresponding Fig. \ref{I-0.3}. The splitting between
the neutrino three point and the others is boosted to about three
hundreths of an eV. 

    Evidently, it may be sometime in the future before the accuracy 
of the beta decay experiments enables one to see all the details
of the fine structure displayed here. The present approach may 
nevertheless be useful in the near future for numerically 
computing the integrated spectrum as a function of both
an energy interval $\Delta$ as well as the running parameter $m_3$
and searching for a best fit to experiment. This would have to be
done for both the type
I and type II cases. A clear best fit would, in principle,
determine all three neutrino masses in a manner consistent with the 
neutrino oscillation data. Of course, the other corrections
mentioned in the Introduction
(which are beyond the scope of the present paper) must be included too.
In the literature \cite{morerefs} there has been some debate about
the proper choice of an effective neutrino
mass to replace all three masses in the analysis of 
endpoint experiments. Clearly the procedure suggested here 
 would enable one to measure
 the effectiveness of different choices. A more detailed discussion
of this aspect will be given elsewhere.

\newpage

\section{Restrictions from a ``complementary" ansatz}
        
    Up to now we have considered only experimental inputs. A key
feature was allowing $m_3$ to vary in order to run through a
family of endpoint spectrum shapes which might be compared with future 
experiments. A similar procedure was recently followed in
order to derive consequences from an ansatz \cite{bfns}-\cite{mns}
which allows one to compute (in case the neutrinos are of Majorana
type) the two Majorana CP violation phases, given $m_3$ and the 
Dirac phase $\delta$. That``complementary ansatz" specifies
that the trace of the prediagonal $3\times 3$ neutrino mass matrix 
vanishes (in a charged lepton diagonal basis wherein non-physical
phases are appropriately chosen). Since the prediagonal neutrino
mass matrix is complex, the vanishing trace gives two physical 
conditions. Taking $A,B,s_{12}^2,s_{23}^2,s_{13}^2$ as ``known",
 running through the two parameters $m_3$ and the Dirac CP
phase $\delta$ then determines the allowed values of the 
two Majorana CP phases. This gives a two parameter family of
solutions for the complete set of neutrino parameters.

    Now in carrying out this analysis, the starting point was
Eqs. (\ref{findmone}) and (\ref{findmtwo}) above. However it 
turns out that the allowed ranges of $m_3$ are more restrictive
than those obtained in and after Eq. (\ref{typeIbound}). In Table 
\ref{m3minvalues}, taken from ref. \cite{mns}, it can be seen that the 
smallest allowed value of $m_3$ for type I solutions is
in the range $0.058-0.059$ eV rather than $0.0517$ eV as found here.
Similarly, the smallest allowed value of $m_3$ for type II solutions
is in the range $0.00068-0.00246$ eV rather than zero as obtained here.
The range of minima exists because there is some dependence on
the input Dirac phase $\delta$.

\begin{table}[htbp]
\begin{center}
\begin{tabular}{lllllll}
\hline \hline type & $(m_3)_{min}$ $(\delta =0)$ in \textrm{ eV} &
 $(m_3)_{min}$ $(\delta =0.5)$
 &  $(m_3)_{min}$ $(\delta =1.0)$ &
  $(m_3)_{min}$ $(\delta =1.5)$&  $(m_3)_{min}$ $(\delta =2.0)$&
$(m_3)_{min}$ $(\delta =2.5)$         \\
\hline
\hline
I &0.0592716 &0.0590967  & 0.0587178 &0.0584799&0.0586203&0.0589971
\\
II & 0.0006811 & 0.00105461 & 0.0019024 &0.0024636&0.0021294&0.0012723    
\\
\hline
\hline
\end{tabular}
\end{center}
\caption[]{Minimum allowed value of the input mass, $m_3$ as
a function of the input CP violation phase, $\delta$ for type
I and II solutions using the ``complementary" ansatz.
  Here, the choice $s_{13}^2 =$ 0.04 has been made.}
\label{m3minvalues}
\end{table}                                                       

\begin{figure}[htbp]
\centering
\rotatebox{0}
{\includegraphics[width=10cm,height=10cm,clip=true]{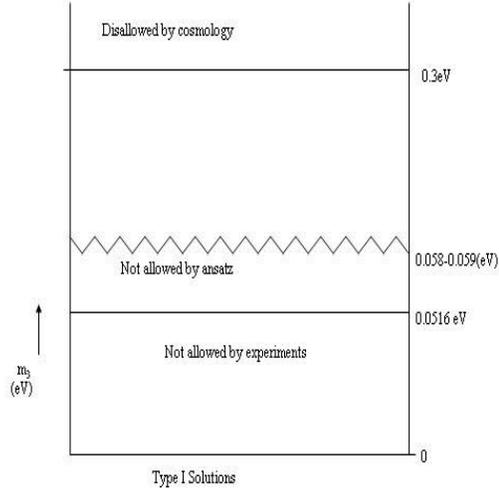}}
\caption[]{Schematic view of the allowed range of neutrino mass $m_3$
for the type I solution.
The wiggly line illustrates the restrictions on the allowed range due to 
assuming the complementary ansatz for the neutrino mass matrix.}
\label{restrictionsI}
\end{figure}

\begin{figure}[htbp]
\centering
\rotatebox{0}
{\includegraphics[width=10cm,height=10cm,clip=true]{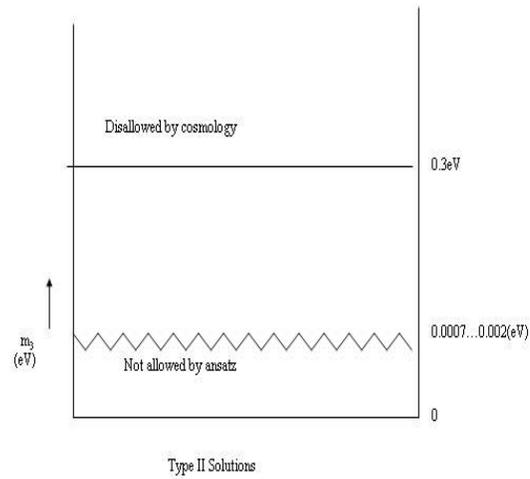}}
\caption[]{Schematic view of the allowed range of neutrino mass $m_3$
for the type II solution. }
\label{restrictionsII}
\end{figure}

    It may be amusing to see why the ansatz
 leads to more restrictions.
Take the simplified case, treated in detail in ref. \cite{nsm},
where $\delta =0$. Then the ansatz condition amounts to 
the lengths corresponding to the three
neutrino masses forming a triangle in the complex plane. (The
two independent Majorana phases are related to the internal angles
of this triangle). The procedure outlined here obtains, for
a given assumed value of $m_3$, the three neutrino masses.
However it is not guaranteed that they form a triangle, since
the mass of one neutrino may be greater than the sum of the 
 masses of the two others. The case where $m_3=0.0517$
in Table \ref{typicalsolutions} is an obvious example of this situation.
 In the case when $\delta$
is non-zero, treated in detail in ref. \cite{mns},
 there is also a triangle condition but the triangle
is not made simply using the masses as sides. At the present moment,
the ``non-triangular" regions of $m_3$ space constitute a small,
but interesting,
part of the allowed range (considering of course the type I and 
type II cases separately). The allowed regions are summarized 
in Figs. \ref{restrictionsI}   and \ref{restrictionsII} .
So, if in the future, the neutrino masses are found
from an endpoint spectrum analysis to give (assuming
suitably updated numbers) $m_3$ outside the range predicted by the
complementary ansatz, the latter can be ruled out.

\section{Summary and discussion}

     We have demonstrated that, given the value of the third neutrino mass
$m_3$, the endpoint spectrum of the electron energy in beta decay is
biuniquely predicted to a reasonable approximation using the experimental 
results
on neutrino oscillations and their analyses. For definiteness some
characteristic endpoint spectrum shapes  were explicitly
illustrated. This seems important not only as an indication of how much our understanding
of neutrinos has advanced in the last ten years but as a convenient
benchmark for helping to analyze future beta decay experiments. Of course
the actual experimental results require other important corrections to this "ideal"
theoretical picture. Some novel features
of this paper which may be of interest include:

    1. While the kinematics of the endpoint spectrum has been discussed
 many times in the literature,
 the exact formula for the phase space dependence on the electron
energy (obtained by performing one integration over the Dalitz diagram while
holding the invariant matrix element constant), Eq. (\ref{phasespace}),
does not seem to have been previously used. These results
may not make big changes in other analyses but one can use them
with confidence since
there will then be no question of kinematical (nucleon recoil) corrections
needing to be separately taken into account. 

    2. The method of analysis of the presently allowed neutrino masses
using the squared mass differences obtained from the neutrino oscillation
 experiments
presented in section III  involves listing the results for the masses of
 neutrinos one and two
while choosing various values of the mass $m_3$. The cases where $m_3$ is
 heaviest (type I)
and where it is lightest (type II) are treated separately. 
 For example, as discussed in section III,
one sees that the currently allowed neutrino mass spectrum is
 characterized for a large part of its range
 as almost degenerate rather than hierarchical. The main point is that for
any choice of $m_3$ and type, the shape of the endpoint spectrum is
 already pretty well known
from the neutrino oscillation experiments. Even though
 the value of $s_{13}^2$ is only
bounded, it 
 will not affect the shape much. Our treatment easily shows
 that the slope discontinuities
corresponding to the mass differences of individual neutrinos occur
 very close to each other.
That would make detailed verification of the shape difficult in
 the near future. However, knowing
the shape for any value of $m_3$ makes it possible to integrate
 the final electron
distribution over an energy interval $\Delta$ corresponding to
 a given experiment. Then the number of
predicted electrons could be plotted as a function of $m_3$ and $\Delta$ 
to get an "ideal" estimate
of the experimental sensitivity to distinguishing different neutrino
 mass scenarios. This ideal
estimate should of course be corrected for effects mentioned in section I, but
that is beyond the scope of the present paper. Further applications
of this approach will be presented elsewhere.   

    3. The complementary ansatz for the neutrino mass matrix
\cite{bfns}-\cite{mns}
 enables one to 
predict, given  $m_3$ and the "Dirac" CP phase, the two Majorana phases of
the lepton mixing matrix. It is of interest because it
 enables the estimation of the neutrinoless
double beta decay quantity $m_{ee}$ and, in a certain model, the baryon asymmetry
 by the leptogenesis mechanism without making arbitrary assumptions about these phases.
 An amusing feature of the ansatz is that (to the
approximation that the effect of the Dirac CP phase $\delta$ is negligible) the
magnitudes of the three neutrino masses must make up a triangle. If they don't,
the ansatz would be falsified. In section IV, the regions of allowed neutrino masses
which falsify the ansatz were specified (including the effect of  $\delta$). It was pointed out that
a large portion of
 the currently allowed range of $m_3$ actually does correspond to a 
triangular pattern
of neutrino masses. However, the small non-triangular region near $m_3
=0.0517$ eV is the only place where $m_3$ is several times higher than
$m_2$.

\section*{Acknowledgments}
\vskip -.5cm
We are grateful to Mariam Tortola for checking our equations
and correcting an overall factor in Eq. (\ref{Edistrib}).
  S.S.M would like to thank the Physics Department at Syracuse University
 for their hospitality.
 The work of S.N is supported by National Science foundation grant No.
PHY-0099.
The work of J.S. is supported in part by the U. S. DOE under
Contract no. DE-FG-02-85ER 40231.

\appendix
\section{Lepton mixing matrix}

    A symmetrical parameterization of the lepton mixing matrix, $K$
can be written as: 
\begin{equation}
K=\left[ \begin{array}{c c c}
c_{12}c_{13}&s_{12}c_{13}e^{i{\phi_{12}}}&s_{13}e^{i{\phi_{13}}}\\
-s_{12}c_{23}e^{-i{\phi_{12}}}-c_{12}s_{13}s_{23}e^{i({\phi_{23}}-{\phi_{13}})}
&c_{12}c_{23}-s_{12}s_{13}s_{23}
e^{i({\phi_{12}}+{\phi_{23}}-{\phi_{13}})}&c_{13}s_{23}e^{i{\phi_{23}}}\\
s_{12}s_{23}e^{-i({\phi_{12}}+{\phi_{23}})}-c_{12}s_{13}c_{23}e^{-i{\phi_{13}}}
&-c_{12}s_{23}e^{-i{\phi_{23}}}-
s_{12}s_{13}c_{23}e^{i({\phi_{12}}-{\phi_{13}})}&c_{13}c_{23}\\
\end{array} \right].
\label{writeout}
\end{equation}
Some advantages of this choice are discussed in sections V
and VI of \cite{mns}. As 
written,
it holds for the case of Majorana type neutrinos. There are three
CP violating phases, $\phi_{12},\phi_{23}$ and $\phi_{13}$.  
 The ``invariant" combination $\delta =\phi_{12} 
+\phi_{23}-\phi_{13}$ corresponds to the ``Dirac phase". If neutrinos
are of Dirac type, only a single phase (say $\phi_{13}$) may be
taken to be non-zero.


\begin{thebibliography}{9}
\bibitem{k}Of course, in recent high accuracy experiments like the
works of \cite{Mainz} and \cite{Troitsk} above, recoil corrections are either
approximated or the end point energy is treated as a fitting parameter.
\end{thebibliography}
\end{document}